\documentclass[11pt,twoside]{article}

\usepackage{asp2006}
\usepackage{epsf}
\usepackage{graphicx}
\usepackage{lscape}
\def\be{\begin{equation}}
\def\ee{\end{equation}}

\def\ergs{{\rm\,erg\,s^{-1}}}
\def\msun{M_{\odot}}

\def\ergs{\rm \,erg\,s^{-1}}
\catcode`\@=11 % This allows us to modify PLAIN macros.
\def\@versim#1#2{\vcenter{\offinterlineskip
        \ialign{$\m@th#1\hfil##\hfil$\crcr#2\crcr\sim\crcr } }}
\def\lsim{\mathrel{\mathpalette\@versim<}}
\def\gsim{\mathrel{\mathpalette\@versim>}}
\def\mpy{M_\odot \ {\rm yr^{-1}}}

\begin{document}

\title{Advection-dominated Accretion: From Sgr A* to other Low-luminosity AGNs}
\author{Feng Yuan}   %%% Fill in author names
\affil{Shanghai Astronomical Observatory, Chinese Academy of
Sciences, 80 Nandan Road, Shanghai 200030, China; and
Joint Institute for Galaxy and Cosmology (JOINGC) of SHAO
and USTC; fyuan@shao.ac.cn}    %%% Fill in author affiliations

\begin{abstract} %%% Abstract to run on from here.

This paper reviews our current understanding of low-luminosity 
AGNs (LLAGNs) in the context of the advection-dominated accretion flow. 
The best investigated source, the supermassive black hole in our
galactic center, Sgr A*, is emphasized since the physics of 
accretion should be the same for various LLAGNs
except for their different accretion rates. 
The important role of jets is discussed, but this is 
less well established.

\end{abstract}
%%% MAIN BODY OF TEXT GOES HERE. CONSULT "INSTRUCTIONS FOR AUTHORS USING
%%% LATEX2E MARKUP", SECTIONS 2.3-2.6 FOR HELP WITH EQUATIONS, FIGURES,
%%% AND TABLES.

\section{Introduction}

It is now widely recognized that most, if not all, galaxies host a
supermassive black hole. Different degree of nuclear activity
are manifested among them, ranging from the most active and luminous 
active galactic nuclei (AGNs), to less active low-luminosity AGNs (LLAGNs),
until the least active quiescent galaxies such as our Galaxy. 
%For the distinctive observational properties of LLAGNs, see \S 3.1 below.
The activities in all these sources are believed 
to be powered by the release of the gravitational energy 
of the gas surrounding the black holes via the accretion process.

The equations of accretion onto a black hole allow two series of solutions,
cool and hot ones. The standard thin disk model (Shakura \& Sunyaev 1973) 
is the representative of the cool solution (another cool accretion solution
corresponding to higher accretion rates is called
``slim'' disk). The temperature of the accretion flow in this solution
is relatively low, $\sim 10^6-10^7$K. It is optically thick, 
geometrically thin (because of the low temperature), and radiatively 
efficient. This solution provides 
a good description to the big blue bump in the optical/UV band of the
luminous AGNs (but see Koratkar \& Blaes 1999) thus is 
believed to work in luminous or normal AGNs. 

The advection-dominated accretion flow (ADAF) belongs to the hot series
(Narayan \& Yi 1994, 1995; Abramowicz et al. 1995;
see Narayan, Mahadevan \& Quataert 1998;
Kato, Fukue \& Mineshige 1998 for reviews). In an ADAF, the temperature 
of ions is virial while the electron temperature is lower but still very high,
$T_e\sim 10^9-10^{11}$K. This solution is 
optically thin and geometrically thick, and most importantly, its radiative 
efficiency is typically much lower than that of the standard thin disk, 
$\eta_{\rm ADAF} \approx 0.1 \dot{M}/\dot{M}_{\rm crit}$. 
Here $\dot{M}_{\rm crit}
\approx \alpha^2\dot{M}_{\rm Edd}$ is the critical accretion rate of ADAF
beyond which the ADAF solution fails and is replaced by another 
hot solution (``luminous hot accretion flow''). There are 
many observational evidence to indicate that ADAFs are very 
likely relevant for understanding LLAGNs. This will be the content of 
this review. 

We will begin our review from Sgr A*, a supermassive black hole
located at the center of our Galaxy. Traditionally our 
Galaxy is regarded as a normal galaxy rather than an AGN. 
However, physically as we will see that
it also manifests some degree of activity, and the exactly same physical
process as in normal AGNs---black hole accretion---is operating there. 
The only difference is that the accretion rate in Sgr A* is much
smaller than in other AGNs and moderately smaller than in LLAGNs. 
So roughly saying, from Sgr A* to LLAGNs to normal AGNs, the 
increasing level of activity is simply because of the increasing 
accretion rate and radiative efficiency. Therefore, 
in this sense Sgr A* is the least luminous LLAGN. We emphasize
Sgr A*, not only because it supplies us with the weakest end of the
activity of AGN population, but as we will see in \S2, it also provides a
unique laboratory of low luminosity accretion.
 
After introducing the ADAF model for Sgr A* in \S2,  
in \S3 we will review our current 
understanding of LLAGNs in the context of ADAFs. Since jets seem to 
be always associated with ADAFs, in \S4 we will discuss the 
 role of jet. 

\section{Sgr A* as a unique laboratory of low luminosity accretion}

\subsection{Why unique? Observational constraints}

The reason why Sgr A* is unique is first because it provides
the best evidence to date for a supermassive black hole (e.g., Sch\"odel 
2002). Secondly, because of its proximity it allows us to observationally 
determine the dynamics of gas quite close to the BH, thus providing uniquely
strict constraints on accretion models.

{\em Outer boundary conditions}. The accretion starts at the Bondi
radius where the gravitational potential energy of the gas equals the 
thermal energy. For uniformly distributed matter with an ambient
density $\rho_0$ and sound speed $c_s$, the Bondi radius
of a BH of mass $M$ is $R_{\rm Bondi}\approx GM/c_s^2$ and the mass
accretion rate is $\dot{M}_{\rm Bondi}\approx 4\pi R_{\rm Bondi}^2\rho_0c_s$.
The high resolution {\em Chandra} observations 
infer gas density and temperature as $\approx 100~ {\rm cm}^{-3}$
and $\approx 2 $ keV on $1^{"}$ scales (Baganoff et al. 2003). The
corresponding Bondi radius $R_{\rm Bondi} \approx 0.04 {\rm pc}\approx
1^{"}\approx 10^5R_s$ and $\dot{M}_{\rm Bondi} \approx 10^{-5} \mpy$. 
The recent 3D numerical simulation for the accretion
of stellar winds on to Sgr A* by Cuadra et al. (2006) obtains
$\dot{M}\approx 3\times 10^{-6}\mpy$, in good agreement with the estimation
of Bondi theory. This simulation also tells us that the angular 
momentum is large, with a circularization radius of about $10^4R_s$.
The Bondi radius, Bondi accretion rate, electrons density and temperature, 
and the angular momentum at the Bondi radius constitute the outer 
boundary conditions any accretion models must satisfy.

{\em Spectral energy distribution}. The spectral
energy distribution of Sgr A* is shown in Fig. 1. The radio
spectrum consists of two component, a low-frequency power-law and 
a ``sub-millimeter bump'', which implies that the radio emission comes from 
two different components. The X-ray emission comes in two states, namely
quiescent and flare ones, with different spectral 
index. A large fraction of the X-ray flux in the quiescent state
comes from an extended region. The bolometric luminosity of Sgr A* 
is only $L\approx 10^{36}\ergs \approx 3\times 10^{-9}L_{\rm Edd}$.
Combined with the Bondi accretion rate, this luminosity implies an 
extremely low radiative efficiency, $\eta\sim 10^{-6}$.

{\em Variability}. At both IR and X-ray wavelengths, the source is 
highly variable. The amplitude of the variability
at IR is $\sim 1-5$ (Genzel et al. 2003;
Ghez et al. 2004); while at X-ray band, it can be as high as 
$\sim 45$ or higher (Baganoff et al. 2001). The variability
at IR and X-ray bands are simultaneous, and the timescales of the variability
are quite similar, typically $\sim$ an hour (Eckart et al. 2006; 
Yusef-Zadeh et al. 2006). 

{\em Polarization}. A high level of linear polarization 
($\sim 2\%-10\%$) at frequencies higher
than $\sim 150$ GHz was detected (e.g., Bower et al. 2003),
which sets an upper limit on the rotation measure of
$7\times 10^5 {\rm rad~m^{-2}}$. This argues for a low density of the
accretion flow at the innermost region of the ADAF. 

\subsection{The standard thin disk is ruled out}

The Bondi theory provides a good estimation to the accretion rate
at the outer boundary. If the gas were accreted at this
rate onto the black hole via the standard thin disk, the expected
luminosity would be $L\approx 0.1\dot{M}_{\rm Bondi} c^2\approx
10^{41}\ergs$, 5~orders of magnitude higher than the observed
luminosity. This is the strongest argument against a standard thin disk in Sgr
A*. In addition, the spectrum shown in Fig. 1 does not look like 
the multitemperature blackbody spectrum predicted by a 
standard thin disk at all.

\subsection{Understanding the observations of Sgr A* with an ADAF}

The crucial point to modeling Sgr A* is that the accretion flow must be
radiatively very inefficient, while the low efficiency is exactly
the characteristic feature of an ADAF solution, as we emphasize in 
the introduction. In a standard thin disk, the viscously dissipated energy 
is radiated away locally, which results in a high efficiency.
But in an ADAF, the radial velocity is much larger 
and the temperature is much higher than
in a standard thin disk. Consequently, the density of the accretion flow 
is much lower. Therefore, the radiative timescale is much longer
than the accretion timescale, thus most of the viscously dissipated
energy is stored in the accretion flow as its thermal energy rather than
radiated away. This is the main reason for the low radiative 
efficiency of an ADAF. For the details of the dynamics of ADAFs, we refer the 
reader to the review of Narayan, Mahadevan \& Quataert (1998). 

\begin{figure}[!ht]
\includegraphics[scale=0.55,angle=0,trim=-50 20 -10 -5]{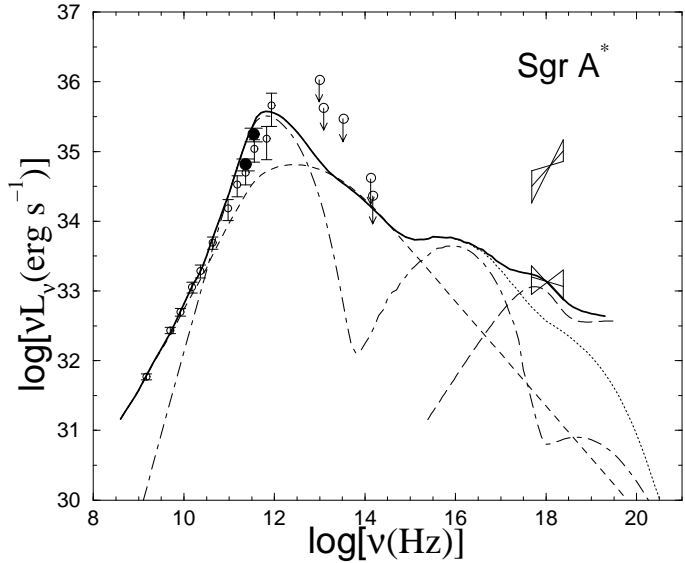}
\caption{ADAF model for the quiescent state emission from
Sgr A*. The dot-dashed line is the
synchrotron and SSC emission by thermal electrons; the dashed line is
the synchrotron emission by non-thermal electrons.
The dotted line is the total synchrotron and SSC emissions while the
solid line also includes the bremsstrahlung emission from the outer parts
of the ADAF (long-dashed line). Adapted from Yuan, Quataert 
\& Narayan (2003).}
\end{figure}

Narayan, Yi \& Mahadevan (1995) first apply the ADAF model to Sgr A*. 
They successfully explain the most important feature of the source,
namely its low radiative efficiency. The spectrum can also be roughly 
explained. However, in the ``old'' ADAF model, the accretion rate 
is assumed to be constant with radius. As a result, the rotation measure
predicted in this model is much larger than that required from the polarization
observation. This is a serious problem of this model. 

In the theoretical side, significant progress has been made in the past decades
in our understanding of ADAFs. First, global, time-dependent,
numerical simulations reveal that only a very small fraction of the
mass that is available at large radii actually accretes onto the black
hole and most of it is lost to a magnetically driven outflow or
circulates in convective motions (Stone, Pringle \& Begelman 1999;
Hawley \& Balbus 2002). Second, in the old ADAF model, the turbulent
dissipation is assumed to heat only ions. However, it was later realized that
processes like magnetic reconnection is likely to heat electrons directly 
(Quataert \& Gruzinov 1999). 

Yuan, Quatatert \& Narayan (2003; 2004) present updated  
ADAF model (also called ``radiatively inefficient accretion flow'')
to Sgr A*, taking into account 
the above-mentioned theoretical developments, i.e., outflow and direct 
electron heating by turbulent dissipation. Fig. 1 shows the spectral
modeling result. Specifically, the submm bump 
comes from the synchrotron emission of thermal electrons in the innermost
region of the ADAF. Due to the existence of outflow, only about 
1\% of the gas available at the Bondi radius enters into the BH horizon,
so the density in the region close to the BH is much lower than in 
Narayan et al. (1995). In this case, the rotation measure is much smaller and
a high linear polarization is expected.
The low-frequency radio and the IR emissions are assumed to be from
some nonthermal electrons in the ADAF (since the plasma in ADAF is 
collisionless). But the low-frequency radio spectrum can 
also be explained by an assumed jet although 
the jet has not been directly detected (e.g., Yuan, Markoff \& Falcke 2002). 
The IR and X-ray flares are explained by the synchrotron and/or SSC emissions
from some transiently accelerated electrons due to processes like magnetic
reconnection in the innermost region of ADAF (Yuan, Quataert \& Narayan 2004).

After the publication of Yuan et al. (2003), some new observations
appeared. A notable one is the size measurement of Sgr A* at radio 
wavebands (Bower et al. 2004; Shen et al. 2005). These results supply 
independent test to theoretical models. Yuan, Shen \& Huang (2006)
calculated the predicted size of Sgr A* by the Yuan et al. (2003) model
and found satisfactory agreement with the observational results. This gives us 
strong confidence for this model, which should be taken as 
a baseline model when modeling other LLAGNs.

\section{Accretion models for other low-luminosity AGNs (LLAGNs)}

\subsection{The main observational results for LLAGNs}

LLAGNs are very common. The Palomar survey (Ho et al. 1997) indicates that 
over 40\% of nearby galaxies contain LLAGNs. They bridge the gap between
the normal galaxies and luminous AGNs. The following 
are the distinctive observational features of LLAGNs (Ho 2005).

\begin{figure}[!h]
%\plotone{llagnsed.eps}
\includegraphics[scale=0.5,angle=-90,trim=60 -20 60 -10]{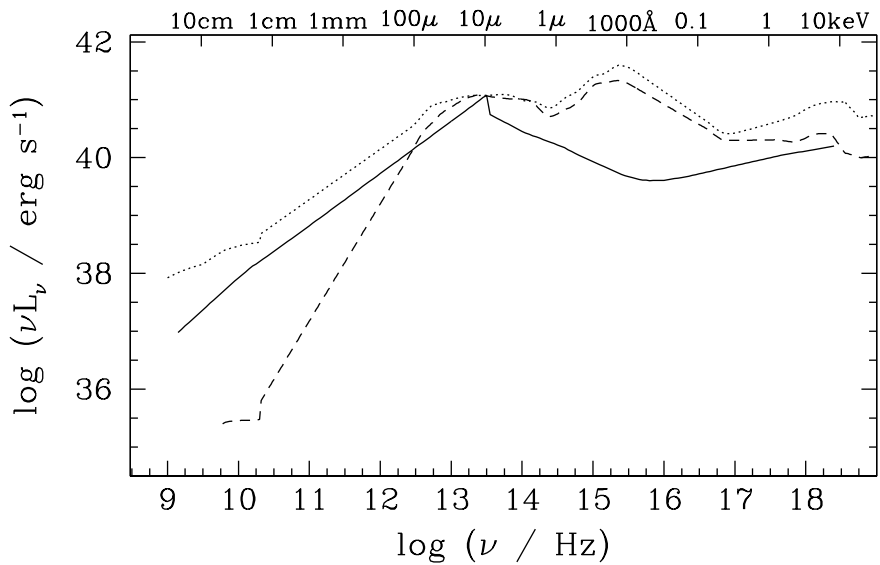}
\caption{The average SED of low-luminosity AGNs (solid line),
compared with the average SEDs of luminous radio-lound (dotted line)
and radio-quiet (dashed line) AGNs. Adapted from Ho (1999).}
\end{figure}

{\em Low accretion power}. LLAGNs are intrinsically faint. Their 
bolometric luminosity $L_{\rm bol}/L_{\rm Edd}\approx 10^{-5}-10^{-3}$,
which is typically lower than Seyferts by one or two orders of magnitude. 
Ho (2005) argues that the low power can not be only explained by the 
low accretion rate and a low radiative efficiency is required. This indicates
that ADAFs must be operating in LLAGNs. 

{\em Unusual spectrum and iron line}. Perhaps the most prominent 
feature of LLAGNs
is the lack of big blue bump in their spectrum. The averaged spectrum of 
LLAGNs is shown by the solid line in Fig. 2. 
Also shown for comparison purpose
is the typical spectra of radio-lound (dotted line) and radio-quiet 
(dashed line) quasars. We see from the figure that in LLAGNs the big blue bump
is missing and instead there is a maximum
peaking somewhere in the mid-IR. The big blue bump is usually associated
with a standard thin disk. Thus the lack of this feature in LLAGNs indicates 
that the thin disk must be absent or truncated. This picture is further 
strengthened by the narrowness of the detected iron K$\alpha$ line. 
This is because if the thin disk had extended to the innermost 
stable circular orbit as in luminous AGNs, the line would be broad. 

{\em Double-peaked Balmer line}. Emission lines with double-peaked
profiles are often found in LLAGNs. Fitting the line profile requires
that the cool accretion disk must have a relatively large inner radius 
(Chen \& Halpern 1989). This is consistent with the result inferred 
from the absences of the big blue bump and broad iron line mentioned above.

\subsection{The current scenario of accretion flow for LLAGNs}

As analyzed above, the accretion picture in LLAGNs is that the thin
disk must be truncated at a certain radius. Within this ``transition''
radius, the thin disk is replaced by an ADAF. A jet is usually required
to fit the radio spectrum of LLAGNs because an ADAF cannot produce enough
radio flux. The possible role of the jet in other wavebands will
be addressed in next section. 

\begin{figure}[!ht]
\includegraphics[scale=0.4,angle=0,trim=-30 -40 10 -50]{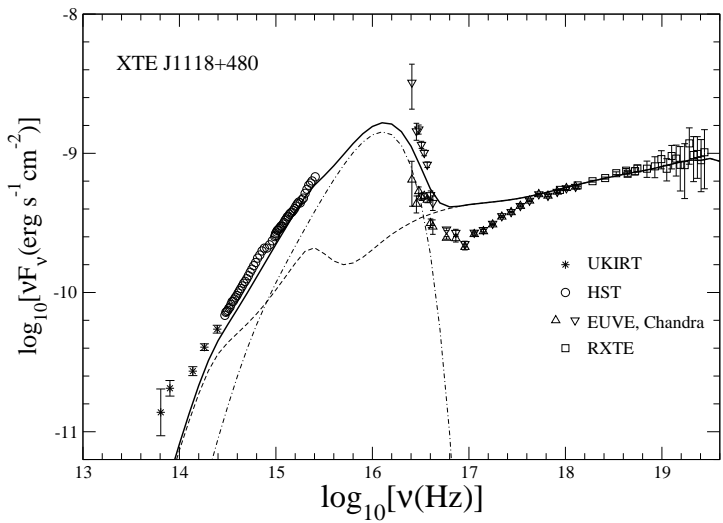}
\caption{ADAF model for the hard state of a black hole X-ray binary---
XTE J1118+480. The dashed, dot-dashed, and solid lines show the emissions
from the ADAF, the truncated thin disk, and their sum, respectively.
Two sets of EUV data correspond to two different choices
of $N_H$. The X-ray spectral break at $\sim 10^{17.7}$Hz is
not real (because of {\em Chandra} calibration issue). Adapted from Yuan, 
Cui \& Narayan (2005).}
\end{figure}

This picture of accretion is originally proposed to explain
the hard state of black hole X-ray binaries (Narayan, McClintock \& Yi 
1996). Since this state is widely believed to 
be the counterpart of LLAGNs, and since so far perhaps the best 
evidence for the truncation of the thin disk still comes from 
one black hole X-ray binary---XTE J1118+480, we first present this example
before we discuss LLAGNs.

The modeling result for the hard state of XTE J1118+480 is shown
in Fig. 3. The optical/UV spectrum is dominated by a truncated
thin disk, while the X-ray emission comes from the Comptonization
of the synchrotron photons in the inner ADAF. The transition radius
is $R_{\rm tr}\approx 300 R_s$. The model
explains the EUV and X-ray data quite well. The under-prediction
of the IR and radio fluxes is because we need to include the contributions
of the jet, as we will discuss in \S4.
Two points need to be emphasized. The first is that there is
a Balmer jump in the optical spectrum and this is believed to be
the evidence for the thin disk origin of the optical emission. The more
important point is on the EUV spectrum which is usually very hard to obtain for
other sources because of the absorption. The high latitude location 
of XTE J1118+480 makes its detection feasible because 
the absorption is very weak. To fit the EUV spectrum, the thin disk must 
be truncated. 

Another important observational result besides the spectrum
is the detection of QPO with frequency of $\sim
0.1$ Hz. One of the most popular models for QPO requires the thin disk must be 
truncated and replaced by an ADAF. The global oscillation of the inner 
ADAF results in the QPO, and the QPO frequency is roughly 
determined by the Keplerian frequency at the transition radius (Giannios \&
Spruit 2004). For XTE J1118+480, the 
Keplerian frequency at $300 R_s$ agrees with the detected QPO frequency
very well within the model uncertainty. So two independent observations,
spectrum and QPO, require the truncation of the thin disk and even give
the same value of the transition radius.

\begin{figure}[!ht]
\plottwo{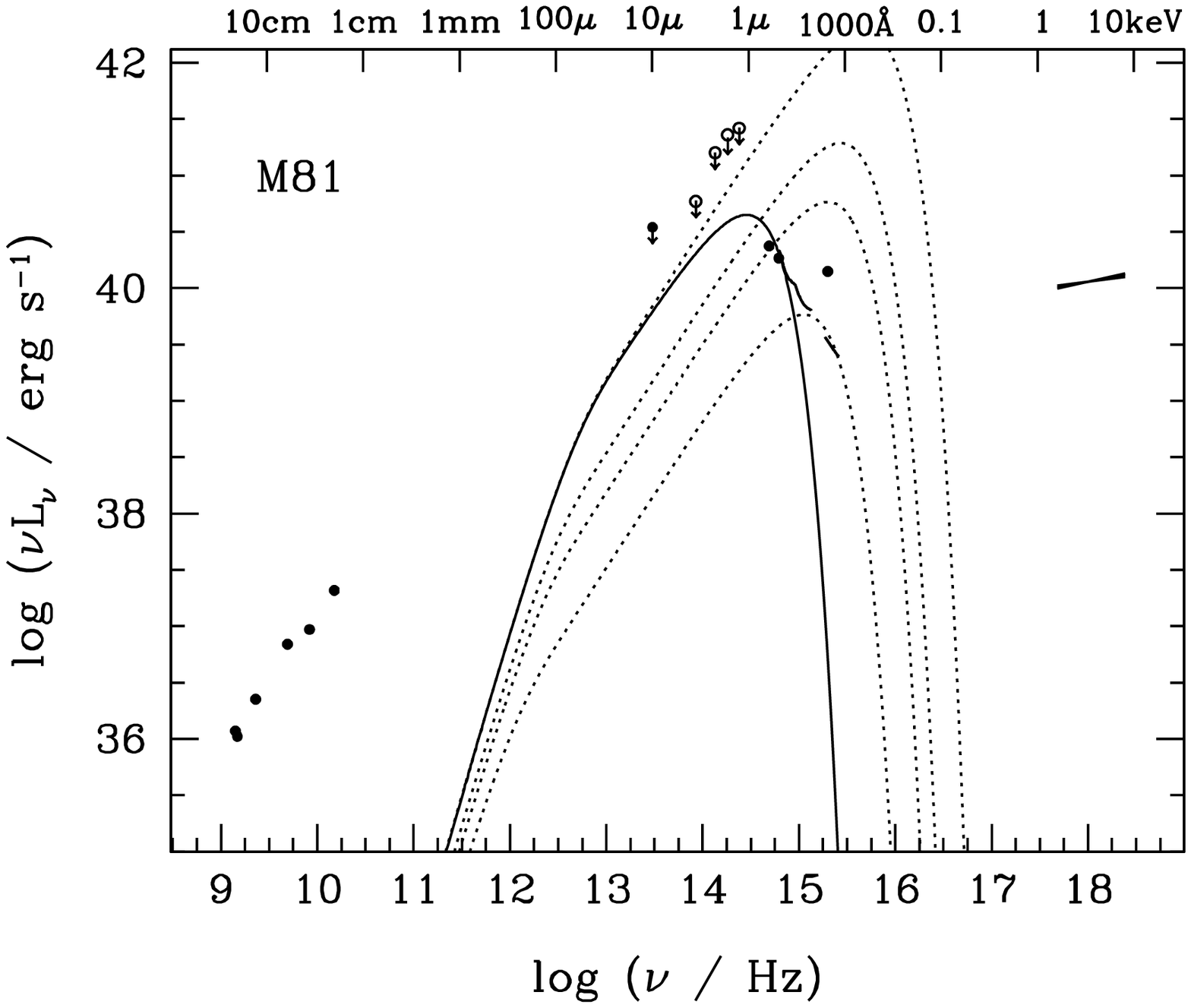}{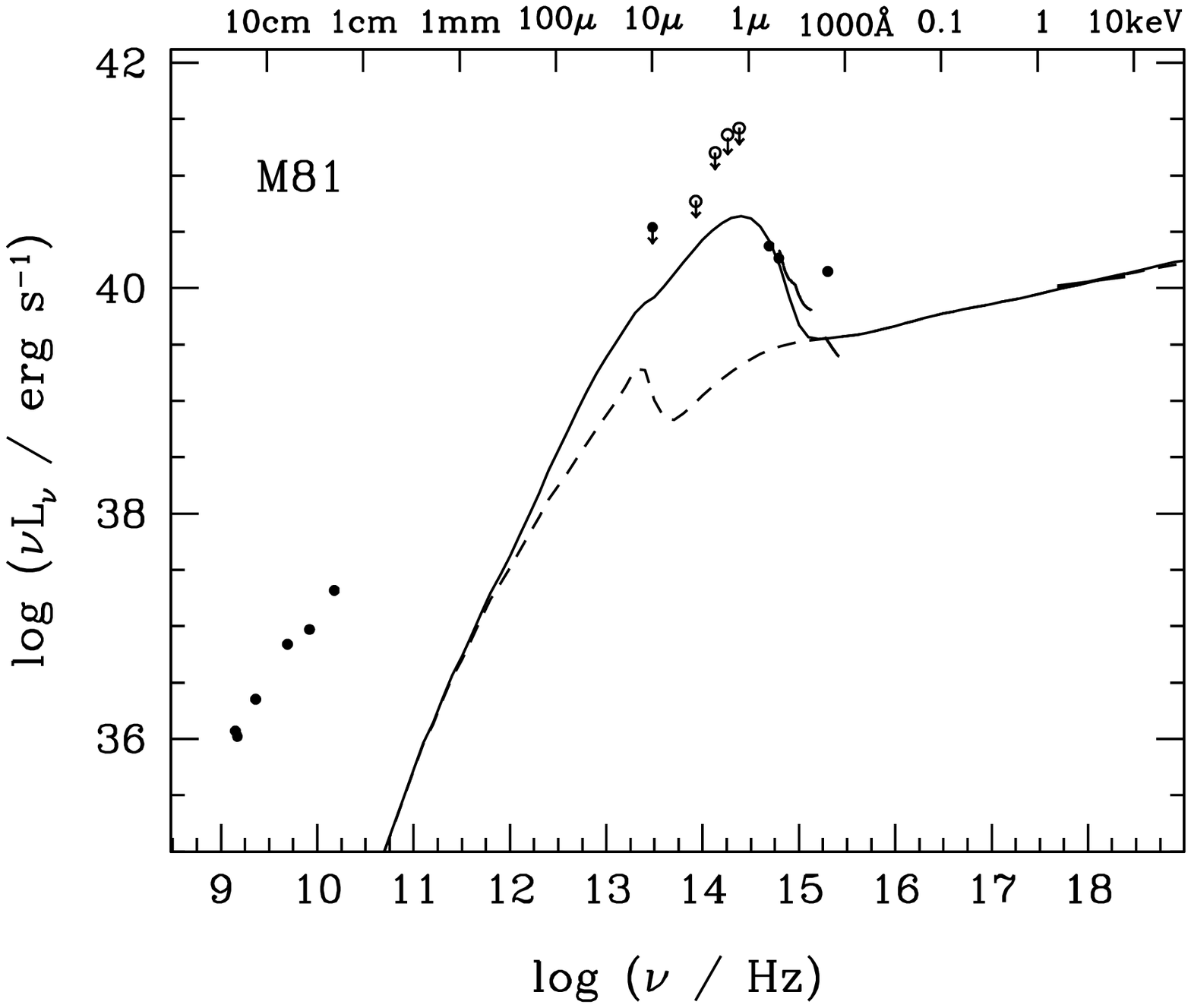}
\caption{Left: The standard thin disk (dotted) and
truncated thin disk (solid) models for the optical/UV spectrum of M 81.
The inner radius of the truncated disk is 100$R_s$. 
Right: The spectral fitting result for a truncated thin disk plus
an inner ADAF. Adapted from Quataert et al. (1999).}
\end{figure}

Fig. 4 shows the modeling result for an LLAGN---M~81 (Quataert et al. 1999). 
We can see that the 
optical/UV spectrum is very steep which is quite different from the 
canonical big blue bump in luminous AGNs. It is clear from the figure that 
such a spectrum cannot be fitted by a standard thin disk without truncation
(shown by the dotted lines in the left panel) but can be fitted 
reasonably well by a truncated disk (shown by the solid line in 
the left panel).

Fig. 5 shows another example of LLAGNs---NGC 1097. This source is 
also well known due to its double-peaked Balmer line. 
It has been known for a long time that to fit such a kind of line profile
the thin disk must be truncated (Chen \& Halpern 1989). 
For NGC 1097, the innermost radius of the thin disk obtained 
is $R_{\rm tr}=252 R_s$ (Storchi-Bergmann et al. 2003). 
On the other hand, to fit the continuum spectrum at optical/UV band, 
we again require that the thin disk must be
required. Although the value of the transition radius is hard to be determined
as precisely as in the case of fitting the line profile, it is shown that 
the value of $R_{\rm tr}=252 R_s$ can fit the optical/UV spectrum 
satisfactorily (Nemmen et al. 2006).
So for this source, like XTE J1118+480,
both two independent observations require a truncated thin disk
and give the same value of the transition radius. 

In addition to the above individual examples, 
ADAF has been proposed to exist in elliptical galaxies (Fabian \& Rees
1995), FR Is (Reynolds et al. 1996; Wu, Yuan \& Cao 2007), XBONGs (Yuan \& 
Narayan 2004), Blazar (Maraschi \& Tavecchio 2003)
and even some Seyfert 1 galaxies (Chiang \& Blaes 2003).
 
\begin{figure}[!ht]
%\plotone{1097.eps}
\includegraphics[scale=0.7,angle=0,trim=-30 -10 10 0]{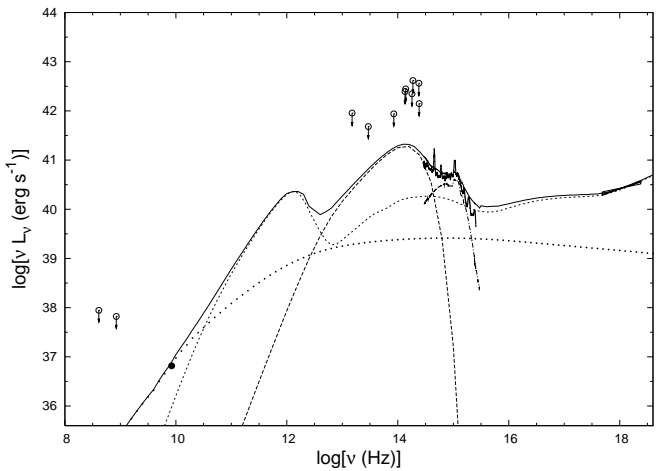}
\caption{Spectral modeling result for NGC 1097. The long-dashed, dotted,
and short-dashed lines show the emitted spectrum by an truncated thin disk,
jet, and an ADAF, respectively. The transition radius is $R_{\rm tr}=225 R_s$.
The dot-dashed line shows the emission from obscured starburst while
the solid shows their sum. Adapted from Nemmen et al. (2006).}
\end{figure}

Several models have been proposed for the mechanism of the 
transition from an outer thin disk to an inner ADAF. Two notable ones are
the evaporation and the turbulent diffusion (e.g., Meyer \& 
Meyer-Hofmeister 1994; Liu et al. 1999; Manmoto \& Kato 2000). 
In the former, the thin disk
is sandwiched by hot corona. The cool matter in the thin disk
will be converted into the hot gas in the corona due to the thermal conduction
between the corona and disk, and finally at a certain radius the whole
thin disk will evaporate thus the transition occurs.
In the latter model, turbulent diffusion can transfer energy from
the inner ADAF to the outer thin disk thus supply energy for the cool thin disk
to make the transition occur. 

%\begin{figure}[!ht]
%\includegraphics[scale=0.6,angle=-90]{4258.eps}
%\caption{Left panel: Snapshot of a star-forming disc of initial mass $2\times
%  10^4$ Solar masses at time $t=10,000$ years. Stars are shown as
%  $t=7,000$ years.}
%%\label{fig:late}
%\end{figure}

\section{Open questions: the role of jets}

It looks like that we now have a good picture of accretion flow
for LLAGNs. However, many details remain to be investigated. 
For example, recently in
the hard state of a couple of black hole X-ray binaries 
some cool material is found to exist very close to 
the innermost stable circular orbit of the black hole. This
seems to imply a standard thin disk without truncation. The iron lines
with broad profile are also claimed to be found in these sources (e.g., Miller
et al. 2006a, 2006b). If confirmed, how to reconcile these results 
within our model will be an interesting and challenging project.

%\begin{figure}[!ht]
%\includegraphics[scale=0.5,angle=0,trim=10 -50 50 0]{corre.eps}
%\caption{The radio--X-ray correlation. From
%  Yuan \& Cui (2005).}
%\label{fig:wind1}
%\end{figure}

Another example of complexity is the possible role of the jet. 
Observationally we know jets are usually associated with ADAFs rather than 
the standard  thin disk, but the physical reason has not been well understood. 
In the context of spectral fitting, it is almost
certain now that the radio emission of most LLAGNs (and hard state of 
black hole X-ray binaries) comes
from the jet rather than the ADAF (e.g., Yuan, Cui \& Narayan 2005). 
But at other wavebands, the role of jets is
still unclear. We can imagine that if the ratio of the mass loss rate
in the jet to the mass accretion rate in the ADAF is high enough (this
could be due to, e.g., a rapidly spinning BH), and if the radiative efficiency
of the jet is high enough (this could be due to the fact
that the shock in the jet is radiative
rather than adiabatic), the spectrum of the source
will be dominated by the jet. This
seems to be the case of NGC~4258 (Yuan et al. 2002).  

An interesting question is that what is the role of jet {\em systematically}
in producing the observed spectrum at other bands? 
For the general sources, it seems unlikely that the jet will dominate
over the ADAF in energy bands above radio, 
say X-ray (e.g., Narayan 2005). Recently
much attention has been paied to the radio--X-ray correlation of BH sources
(e.g., Merloni et al. 2003). Note this correlation does not mean
that the X-ray and radio emissions must have the same origin---jets.
This is because if the accretion rate in the ADAF is positively correlated 
with the mass loss rate in the jet, which 
is very likely, we would naturally expect such a correlation. 

Based on the Merloni et al. (2003) correlation 
and some assumptions to the physics of jets,
Yuan \& Cui (2005) argue that in the X-ray band
the (synchrotron) emission of the jet may dominate over the underlying ADAF,
when the X-ray luminosity of the system is below a critical value determined by:
\be
{\rm log}\left(\frac{L_{\rm x,crit}}{L_{\rm Edd}}\right)
=-5.356-0.17{\rm log}\left(\frac
{M}{\msun}\right).
\ee
The modeling results to a sample of 
FR Is seems to support this prediction (Wu, Yuan \& Cao 2007;
see also other possible observational evidences presented 
in Yuan \& Cui 2005). The sample adopted only consists of
eight sources which is too few. A much larger sample and better data are 
required to examine this prediction. 

\acknowledgements I think the organizers of the conference for 
their invitation. This work is supported in part by Bairen Program of CAS.

%%% THE BIBLIOGRAPHY
%%%
%%% CONSULT SECTION 3 OF "INSTRUCTIONS FOR AUTHORS" FOR HOW TO USE NATBIB.
%%% AUTHORS ARE ENCOURAGED TO USE EITHER THE "THEBIBLIOGRAPY" ENVIRONMENT
%%% BY UNCOMMENTING (DELETING THE "%" SYMBOL) THE COMMANDS BELOW, OR BY
%%% USING THE BIBTEX ENVIRONMENT. TO FIND OUT WHICH IS APPLICABLE TO YOUR
%%% CONTRIBUTION, CONSULT THE VOLUME EDITORS FOR YOUR PROCEEDINGS.
%%%

\end{document}